\documentclass[preprint,sort&compress, 5p,times]{elsarticle}

\usepackage{lipsum}
\makeatletter
\def\ps@pprintTitle{%
 \let\@oddhead\@empty
 \let\@evenhead\@empty
 \def\@oddfoot{}%
 \let\@evenfoot\@oddfoot}
\makeatother

\usepackage{graphicx}
\usepackage{amssymb}
\usepackage{amsmath}
\DeclareMathOperator{\csch}{csch}
\DeclareMathOperator{\Kn}{Kn}
\newcommand{\EL}{\ensuremath{\mathcal{L}}}
\newcommand{\PL}{\ensuremath{\mathcal{P}_L}}
\newcommand{\PT}{\ensuremath{\mathcal{P}_T}}

\journal{}

\begin{document}

\begin{frontmatter}


\title{Hydrodynamic attractors for Gubser flow}



\author{Ashutosh Dash}
\ead{ashutosh.dash@niser.ac.in}
\author{Victor Roy}
\ead{victor@niser.ac.in}
\address{National Institute of Science Education and Research, HBNI, 752050 Odisha, India.}

\begin{abstract}
The Boltzmann equation is solved in the relaxation time approximation using a hierarchy of angular moments of the distribution function.
Our solution is obtained for an azimuthally symmetric radially expanding boost-invariant conformal system that is undergoing Gubser flow. 
The solution of moments that we get after truncating the infinite set of equations at various orders is compared to the exact kinetic solution. 
The dynamics of transition is described by the presence of fixed points which describes the evolution of the system from an early time collisionless free streaming to the hydrodynamic regime at intermediate times and back to free streaming at late times. The attractor solution is found for various orders of moments as an interpolation between these fixed points. The relation of moments to various approximations of relativistic viscous hydrodynamics is investigated. 
\end{abstract}


\end{frontmatter}


\section*{Introduction}
\label{S:1}
Hydrodynamics is an effective theory for the description of long-wavelength phenomena of fluids, that can be expressed as a  
small gradient expansion relative to a thermal background \cite{Florkowski:2017olj}. Thus, hydrodynamics is expected to fail for
systems which are far-from-equilibrium where the gradients are expected to be large. The medium produced in $pp$ collisions at LHC and RHIC
energies is an example of such a system. However, recent experimental results of high energy $pp$ collision have shown evidence of collectivity similar
to those observed in heavy-ion collisions \cite{Khachatryan:2016txc,Weller:2017tsr,Bozek:2010pb,Werner:2010ss,Romatschke:2016hle}. The unprecedented 
success of hydrodynamics to describe collectivity in heavy-ion collisions, as well as small systems, can be attributed to the fact that there exists a
stable universal attractor which makes the dynamical equations to quickly converge and enter a hydrodynamic regime, at a time scale much smaller than the
typical isotropization time scales \cite{Heller:2014wfa, Heller:2016rtz, Denicol:2016bjh, Bazow:2016oky, Heller:2015dha,Giacalone:2019ldn,Kurkela:2019set}.

Previous works \cite{Heller:2016rtz, Aniceto:2015mto, Strickland:2017kux, Florkowski:2017jnz, Strickland:2018ayk, Jaiswal:2019cju, Chattopadhyay:2019jqj, Behtash:2019txb} have mostly focused
on studying the properties of attractors for rapidly expanding 1+1d boost invariant systems undergoing Bjorken flow using relativistic kinetic theory. However, 
the fireball produced in high energy heavy-ion collisions also expands in the transverse direction at late times. Therefore, it is natural to ask whether the systems undergoing simultaneous longitudinal and transverse expansion also shows the attractor nature as observed for 1+1d expansion.  In the present work, 
we consider a system undergoing Gubser flow which has a simultaneous transverse and longitudinal expansion.  In the context of Gubser flow, some recent works \cite{Behtash:2017wqg, Denicol:2018pak} study such dynamical properties using anisotropic hydrodynamics or DNMR hydrodynamical theories \cite{Denicol:2012cn}. Recently in \cite{Behtash:2019qtk}, it was found that the Gubser dynamical system allows a perturbative series solution around its early and late time fixed points with a  finite radius of convergence. Based on these findings, it was shown that Gubser expanding system does not hydrodynamize. The purpose of the present work is to investigate the dynamics of transition from free streaming regime to a hydrodynamic regime or vice versa using a moment method that translates the kinetic equation for distribution function in relaxation time approximation (RTA) to an infinite series of coupled ordinary differential equations advocated in
\cite{Blaizot:2017lht, Blaizot:2017ucy, Blaizot:2019scw}. It will be seen that unlike 1+1d Bjorken flow which has late-time thermalization/hydrodynamization, Gubser expansion is intrinsically 3+1d expansion with dynamics such that the system goes from early time free-streaming regime to intermediate thermalization/hydrodynamization and back to free-streaming in the late time regime.

\section*{Formulation}
The hydrodynamic equations can be derived from the transport equation by taking appropriate moments of the distribution function.
The collision term in the Boltzmann equation plays an important role to isotropize any arbitrary initially anisotropic out of equilibrium distribution function or the
anisotropy generated by a strong expansion. The competition between the two effects is commonly investigated in terms of the ratio between longitudinal pressure \(\mathcal{P}_{L}\) and transverse  \(\mathcal{P}_{T}\), local equilibrium corresponds to   \({\mathcal{P}_{T}}/ { \mathcal{P}_{L} }=1\) . 
In \cite{Blaizot:2017lht} it has been shown that a particular moment of the distribution function is very useful for studying an out of equilibrium system undergoing longitudinally boost-invariant expansion. The details of the moment of the distribution function will be discussed later. 
The Boltzmann equation in an arbitrary coordinate system for on-shell particles is
\begin{equation}
\label{eq:BoltzGen}
p^{\mu} \partial_{\mu} f+\Gamma_{\mu j}^{\lambda} p_{\lambda} p^{\mu} \frac{\partial f}{\partial p_{j}}=\mathcal{C}[f],
\end{equation}

where \(\Gamma_{\mu j}^{\lambda}\) are the Christoffel symbols and \(f=f(x^{\mu},p^{j})\) is the one particle distribution function. 

Following \cite{Gubser:2010ze,Gubser:2010ui} we look for solutions of the hydrodynamic equations with $SO(3)_{q} \otimes SO(1,1) \otimes Z_{2}$ symmetry in flat spacetime. 
This can by implemented by a Weyl transformation of Minkowski spacetime in Milne coordinates $x^{\mu}=(\tau, x, y, \eta)$ to $d S_{3} \otimes \mathbb{R}$ 
spacetime. Here 3 stands for the 3-dimensional de Sitter spacetime assuming that the fluid is homogeneous. This spacetime is described by the line element
\begin{equation}
 d \hat{s}^{2}=-d \rho^{2}+\cosh ^{2} \rho\left(d \theta^{2}+\sin ^{2} \theta d \phi^{2}\right)+d \eta^{2}
\end{equation}
Here we have introduced the de Sitter coordinates $\hat x^{\mu}=(\rho,\theta,\phi,\eta)$, with
\begin{eqnarray}
&\rho(\tau, r)=-\sinh ^{-1}\left(\frac{1-q^{2} \tau^{2}+q^{2} r^{2}}{2 q \tau}\right),\\
&\theta(\tau, r)=\tan ^{-1}\left(\frac{2 q r}{1+q^{2} \tau^{2}-q^{2} r^{2}}\right).
\end{eqnarray}
$q^{-1}$ is an arbitrary length scale and sets the size of the system. In these coordinates, the Gubser flow profile
takes the form: $u_{\tau}=-\cosh \kappa(\tau, r), u_{r}=\sinh \kappa(\tau, r),$ with transverse flow
rapidity $\kappa=\tanh ^{-1}\left(2 q^{2} \tau r /\left(1+q^{2} \tau^{2}+q^{2} r^{2}\right)\right)$. 
In this paper, all quantities in de Sitter coordinates are denoted with a hat.

Gubser symmetry requires massless degrees of freedom of the fluid, i.e., ${\hat p}^2=0$ and thus the spatial momentum can be written as
\begin{equation}
 \hat{p}_{\rho}=\sqrt{\left(\hat{p}_{\theta} / \cosh \rho\right)^{2}+\left(\hat{p}_{\phi} /(\cosh \rho \sin \theta)\right)^{2}+\hat{p}_{\eta}^{2}}.
\end{equation}
The ${SO(3)}_{q}$ symmetry implies the distribution function to be independent of $(\theta,\phi)$ and that the distribution function depends only on the following
combination of momentum components:
\begin{equation}
 \hat{p}_{\Omega}^{2}=\hat{p}_{\theta}^{2}+\frac{\hat{p}_{\phi}^{2}}{\sin ^{2} \theta}.
\end{equation}
Similarly the $SO(1,1)$ symmetry subgroup implements longitudinal boost invariance as a result of which the distribution function is independent of $\eta$.
Using the above constraints, the RTA Boltzmann equation Eq.~(\ref{eq:BoltzGen}) in de Sitter space takes the following form \cite{Denicol:2014xca,Denicol:2014tha}
\begin{equation}\label{Eq:Boltzmann}
\frac{\partial}{\partial \rho} f\left(\rho ; \hat{p}_{\Omega}, \hat{p}_{\eta}\right)=-\frac{\hat{T}(\rho)}{c}\left[f\left(\rho ; \hat{p}_{\Omega}, 
\hat{p}_{\eta}\right)-f_{\mathrm{eq}}\left(\hat{p}^{\rho} / \hat{T}(\rho)\right)\right].
\end{equation}
Here $\hat{p}^{\rho}=\sqrt{\left(\hat{p}_{\Omega} / \cosh \rho\right)^{2}+\hat{p}_{\eta}^{2}}$ and  $\hat{T}=\tau T$. 
$c$ is a  dimensionless parameter which, in RTA,  can be expressed in terms of the shear viscosity $(\eta_s )$ to entropy density $(s)$ ratio $\bar{\eta}_s \equiv \eta_s / s$ 
as $c=5 \bar{\eta}_s$.

\section*{Evolution of moments in Gubser flow}
Although Eq.~(\ref{Eq:Boltzmann}) can be solved numerically, much insight can be gained by taking various moments of the distribution function
as elucidated in \cite{Blaizot:2017lht,Blaizot:2017ucy}. These moments capture the deviation of distribution function from isotropy and for an
expanding system they approach local equilibrium at late times due to collisions. The $n^{th}$ order moment 
$\mathcal{L}_n$ of the distribution  function is defined as :
\begin{equation}\label{Eq:Moments}
 \EL_n=\int d \hat{P}\left(\hat{p}^{\rho}\right)^{2}P_{2n}\left(\hat{p}_{\eta}/\hat{p}^{\rho}\right) f\left(\rho ; \hat{p}_{\Omega}, \hat{p}_{\eta}\right),
\end{equation}
where $d \hat{P}=d \hat{p}_{\eta} d \hat{p}_{\theta} d \hat{p}_{\phi} /\left((2 \pi)^{3} \hat{p}^{\rho} \cosh ^{2} \rho \sin \theta\right)$ is the phase space measure
and $P_{2 n}$ is a Legendre polynomial of order $2 n$. Except for the moment $n=0$ which corresponds to energy density $\hat{\epsilon}$, all higher order $n$ quantify the details of 
longitudinal momentum anisotropy. For example, the first order moment $\EL_1=\hat{\PL}-\hat{\PT}$, describes the anisotropy of pressure components in longitudinal and transverse 
directions. 

\begin{figure}
    \centering
    \includegraphics[height =0.6\textwidth]{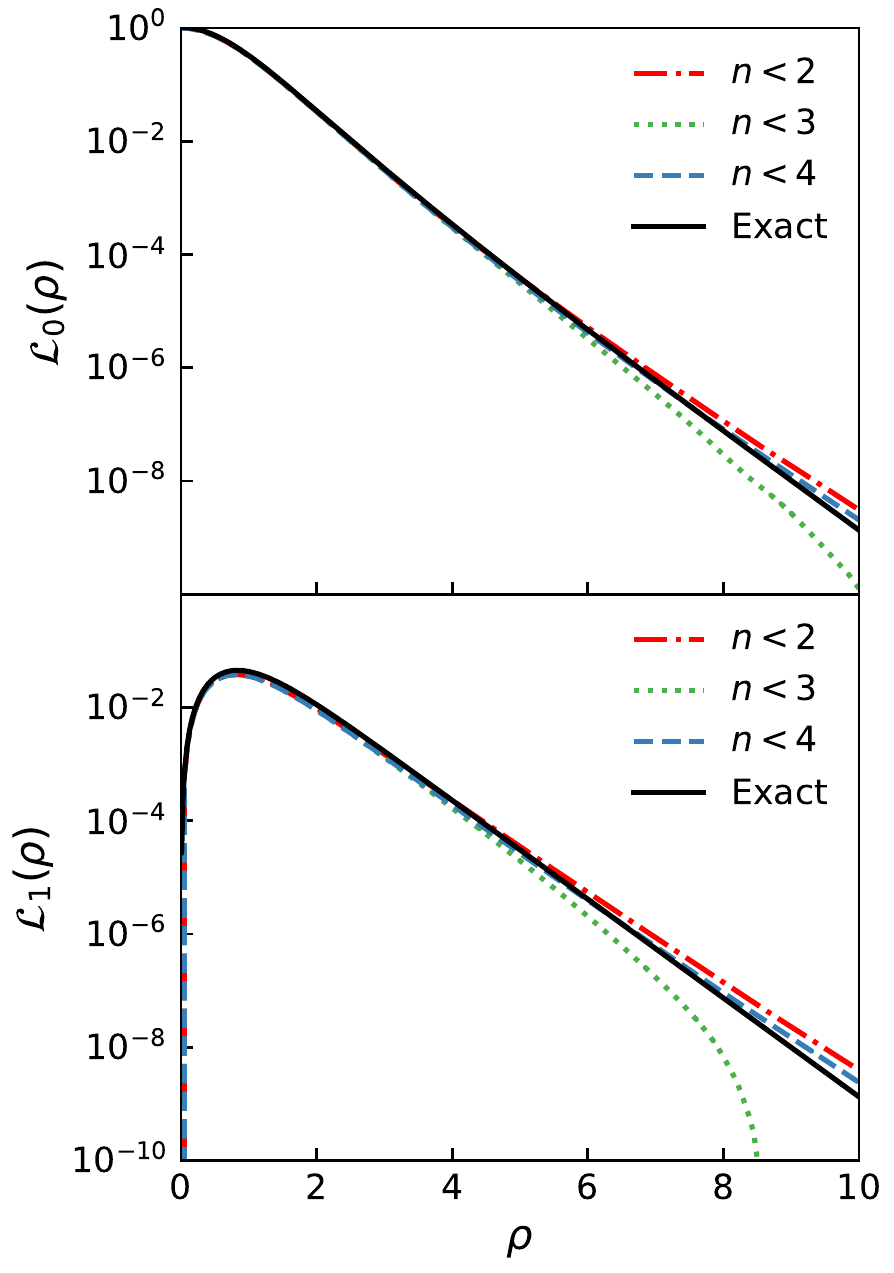}
     \caption{(Color online) (a) $\mathcal{L}_0$ moment by truncating Eq.~(\ref{Eq:Main}) at various orders. The initial condition is that of an equilibrium distribution
      function imposed at $\rho_0=0$ with $\mathcal{L}_{0}(\rho_0) =1$. The results are obtained assuming $\bar \eta=1/4\pi$. (b) Same as (a) but for $\mathcal{L}_1$.}
    \label{fig:Compare}
\end{figure}
Taking the derivative of Eq.~(\ref{Eq:Moments}) with respect to de Sitter time $\rho$ and using Eq.~(\ref{Eq:Boltzmann}) one arrives at the following infinite set 
of coupled differential equations,
\begin{align}
 \frac{\partial \EL_n}{\partial \rho}&=-\tanh(\rho)\left(a_n \EL_n + b_n \EL_{n-1} + c_n \EL_{n+1}\right)-\hat{T}(\rho)\frac{\EL_n}{c}, & (n\geq1)\nonumber \\
 \frac{\partial \EL_0}{\partial \rho}&=-\tanh(\rho)\left(a_0 \EL_0 + c_0 \EL_{1}\right), & (n=0)\label{Eq:Main}
\end{align}
where the coefficients $a_n, b_n$, and $c_n$ are given as
\begin{eqnarray}\label{Eq:coeff}
 a_n&=&\frac{2 (18 n^2+9 n-4)}{(4 n-1) (4 n+3)},\\
 b_n&=& -\frac{4 n (n+1) (2 n-1)}{(4 n-1) (4 n+1)},\\
 c_n&=&\frac{2 (n+1) (2 n-1) (2 n+1)}{(4 n+1) (4 n+3)}.
\end{eqnarray}
As one can clearly see from Eq.~(\ref{Eq:Main}),  there is a competition between collisions (the term containing relaxation time) which washes
out the effect of anisotropy and expansion (determined from the dimensionless coefficients $a_n, b_n$, and $c_n$) which drive the system out of
equilibrium. One should note that a key difference between the present work and that
of \cite{Blaizot:2017ucy} is that the relaxation time $\tau_R$ in a conformal setting is not a constant but is related to the only available 
scale in the system i.e., temperature through the relation
$\hat{\tau}_R=c/\hat{T}(\rho)$. A comparison of the coefficients $a_n$, $b_n$, and $c_n$ for Gubser flow against the coefficients $\tilde{a}_n$,
$\tilde{b}_n$. and $\tilde{c}_n$ for Bjorken flow of \cite{Blaizot:2017ucy} gives us the following relations: $a_n+\tilde{a}_n=4$, $b_n=-\tilde{b}_n$
and $c_n=-\tilde{c}_n$. \par

The system of equations Eq.~(\ref{Eq:Main}) can be easily solved by truncating the series at a given $n$ and ignoring all higher-order terms. In Fig.~\ref{fig:Compare}
we compare the solution of these equations with the exact numerical solution of Eq.~(\ref{Eq:Boltzmann}) by using a method given in \cite{Denicol:2014tha}. 
First, we see that even for $n<2$, the solution captures the qualitative details of the exact solution. Second, one notices that the approach to the exact solution
is alternating for even and odd $n$. We would like to also point that, the exact kinetic solution shows numerical instability at negative $\rho$ values when the system
is initialized with non-zero shear stresses at $\rho_0=0$. The same problem also persists in the moment method when we try to extrapolate the solution to negative de
Sitter coordinates.

\section*{The free streaming fixed point}
\begin{figure}
    \includegraphics[width =0.45\linewidth]{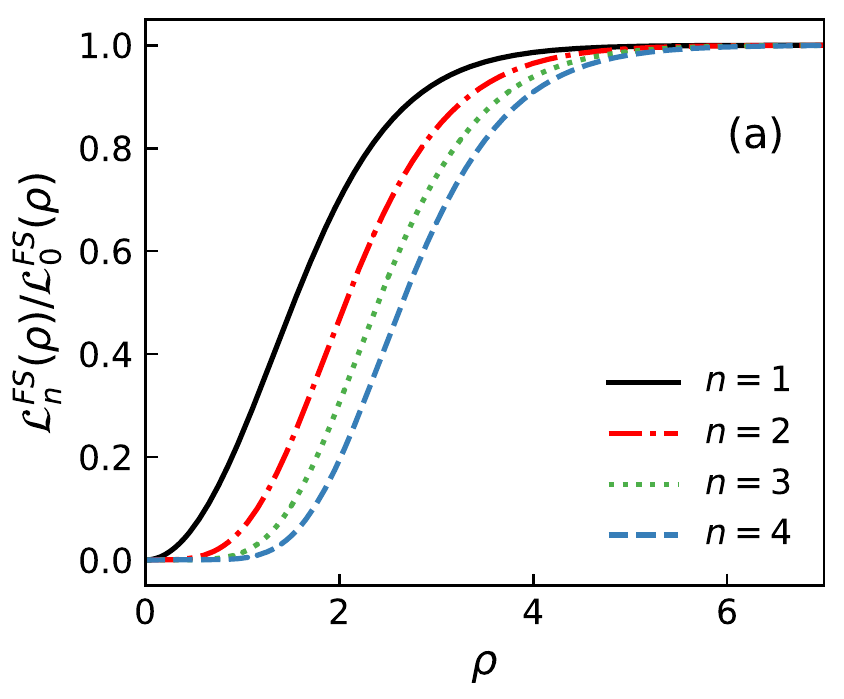}
     \includegraphics[width =0.45\linewidth]{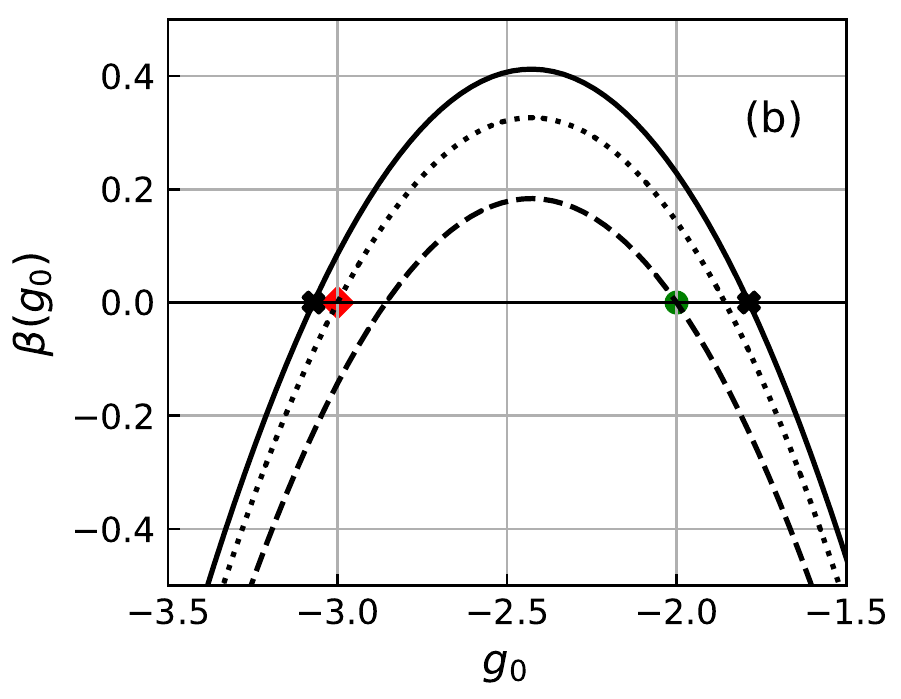}
     \caption{(Color online) (a) Scaled $\mathcal{L}_n$ moments in the free streaming limit, with equilibrium initial conditions imposed at 
     $\rho_0=0$ with $\hat{T}(\rho_0) =1$, obtained from Eq.~(\ref{Eq:FS}). (b)~$\beta(g_0)$ as a function of $g_0$ from Eq.~(\ref{Eq:betag0}). The crosses indicate the approximate solutions obtained from Eq.(\ref{Eq:betag0}). The red diamond point corresponds to the late time unstable fixed point which is obtained from   Eq.(\ref{Eq:betag0Exact}) for $\frac{\EL_2}{\EL_0}=\frac{3}{8}$.  The green circle corresponds to a late time stable fixed point which is obtained from   Eq.(\ref{Eq:betag0Exact}) for $\frac{\EL_2}{\EL_0}=1$ .}  . 
    \label{fig:FS}
\end{figure}
The RTA Boltzmann equation Eq.~(\ref{Eq:Boltzmann}) has the following exact solution \cite{Denicol:2014xca,Denicol:2014tha}:
\begin{align}\label{Eq:ExactBoltz}
f\left(\rho ; \hat{p}_{\Omega}^{2}, \hat{p}_{\eta}\right)&=D\left(\rho, \rho_{0}\right) f_{0}\left(\rho_{0} ; \hat{p}_{\Omega}^{2}, \hat{p}_{\eta}\right)+\\
&\ \frac{1}{c}\int_{\rho_{0}}^{\rho} d \rho^{\prime} D\left(\rho, \rho^{\prime}\right) \hat{T}\left(\rho^{\prime}\right) f_{\mathrm{eq}}\left(\rho^{\prime} ; \hat{p}_{\Omega}^{2}, 
\hat{p}_{\eta}\right),
\end{align}
where $D\left(\rho, \rho_{0}\right)=\exp \left[-\int_{\rho_{0}}^{\rho} d \rho^{\prime} \hat{T}\left(\rho^{\prime}\right) / c\right]$ is 
the damping function, $f_{0}\left(\rho_{0} ; \hat{p}_{\Omega}^{2}, p_{\eta}\right)$ is the initial distribution function at de Sitter time $\rho_{0}$ and
$f_{\mathrm{eq}}\left(\rho^{\prime} ; \hat{p}_{\Omega}^{2}, \hat{p}_{\eta}\right)$ is the equilibrium distribution function. The moments can be calculated 
from the distribution function Eq.~(\ref{Eq:ExactBoltz}) using the definition given in Eq.~(\ref{Eq:Moments}).

Assuming that the initial distribution $f_{0}\left(\rho_{0} ; \hat{p}_{\Omega}^{2}, \hat{p}_{\eta}\right)$ at $\rho_{0} $ is an isotropic Boltzmann distribution,
 we find the following free streaming (FS) solution (taking $c\rightarrow \infty$ in Eq.~(\ref{Eq:ExactBoltz})) for the moments 
\begin{equation}\label{Eq:FS}
 \EL^{FS}_n(\rho)=\frac{3 \hat{T_0^{4}}}{\pi^2}x^2\mathcal{F}_{n}\left(x\right),
\end{equation}
where $x=\frac{\cosh\rho_0}{\cosh\rho}$ and $\mathcal{F}_{n}(x)$ as
\begin{equation}\label{Eq:Fnx}
\mathcal{F}_n(x)= \frac{1}{2}\int_{-1}^{1}dy~\sqrt{\left(1-x^2\right) y^2+x^2} P_{2 n}\left(\frac{y}{\sqrt{x^2+\left(1-x^2\right) y^2}}\right).
\end{equation}
Here $\hat{T_0}$ is the initial temperature at $\rho_{0}$. The function $\mathcal{F}_n(x)$ has the following
limits: $\mathcal{F}_{n}(x)\rightarrow 1/2$ as $x\rightarrow 0$ and $\mathcal{F}_{n}(x)\rightarrow \frac{\sin 2n\pi}{2n\pi\left(1+2n\right)}$ as $x\rightarrow 1$
(i.e, $\mathcal{F}_0=1$ for $n=0$ and $\mathcal{F}_0=0$ for $n\neq 0$). 
Consequently, for asymptotically large de Sitter time, i.e., $|\rho|\gg|\rho_0|$, $\mathcal{F}_{n}(x)$ is a constant and $\mathcal{L}^{FS}_n(\rho)$
decays as $1/{\cosh^2\rho}$. However, the scaled moments,
\begin{equation}\label{Eq:StableFS}
 \lim_{\rho\rightarrow\infty}\frac{ \EL^{FS}_n(\rho)}{\EL^{FS}_0(\rho)}=1,
\end{equation}
saturate in the FS limit, which would have decayed to zero in the presence of collisions.

In Fig.~\ref{fig:FS} we show the analytical solution of the evolution of scaled moments by solving Eq.~(\ref{Eq:FS}). As one can see, all the scaled moments approach unity
at large de Sitter times. The FS regime can also be described by Eq.~(\ref{Eq:Main}) by taking the limit $c\rightarrow \infty$. We have also checked 
(not shown here) that even
for arbitrary initial condition, the moments approach unity both from above and below at large de Sitter times. \par

Since the moments continuously evolve to a constant value at large times, it is natural to define the quantity $g_n(\rho)$ \cite{Blaizot:2017lht,Blaizot:2017ucy},
\begin{equation}\label{Eq:gnDef}
 g_n(\rho)=\frac{\partial\ln \EL_n}{\partial\ln(\cosh \rho)}.
\end{equation}
For Eq.~(\ref{Eq:Main}), $g_n(\rho)$ turns out to be,
\begin{equation}\label{Eq:FP}
 g_n(\rho)=-a_n - b_n \frac{\EL_{n-1}}{\EL_n} - c_n \frac{\EL_{n+1}}{\EL_n}-(1-\delta_{n0})\frac{\hat{T}(\rho)}{c\tanh\rho}.
\end{equation}
Taking the limit $c\rightarrow \infty$, and using the expression Eq.~(\ref{Eq:StableFS}) for the ratio of the moments (for large \(\rho\))
 one can obtain the FS fixed point. We find that for all $n$ the solution yields $g_n(\rho)=-2$. 
For the above initial condition, one finds that solution of $g_n(\rho)$ from Eq.~(\ref{Eq:Main}) does not evolve with $\rho$ and hence is indeed
a fixed point. One may also verify that for arbitrary initial conditions, the series of equations Eq.~(\ref{Eq:Main}) also gives the approximate
result $g_n(\rho)\approx-2$ with error due to finite truncation of the series. Alternatively, one may define an equation of
motion for the quantity $g_n(\rho)$,
\begin{equation}\label{Eq:betaEvoln}
\frac{\partial g_n}{\partial\ln(\cosh \rho)}=\beta(g_n).
\end{equation}
The fixed points correspond to the zeros of $\beta(g_n)$. For $n=0$ and assuming vanishing $\EL_2$ and higher order moments, $\beta(g_0)$ turns out to be
\begin{equation}\label{Eq:betag0}
 \beta(g_0)\approx-g_0^2-(a_0+a_1)g_0+c_0b_1-a_0a_1.
\end{equation}
The plot of the function $\beta(g_0)$ is shown in Fig.~\ref{fig:FS}(b). We find that there is yet another fixed point $g_0\approx-3$ apart from the 
previously found point $g_0\approx-2$. Applying linear perturbation around the fixed point $\langle g_0\rangle$, i.e., $g_0(\rho)=\langle g_0\rangle+f(\rho)$,
and substituting this in Eq.~(\ref{Eq:betaEvoln}) we find that the solution around $\langle g_0\rangle=-2$ yields decaying modes for $\rho>0$ and hence a stable fixed
point in the late time regime where as the solution at $\langle g_0\rangle=-3$ yields growing modes for $\rho>0$ and therefore an unstable fixed point in the late time regime. However, in the early time regime $\rho<0$, $g_n=-3$ is stable and $g_n=-2$ is unstable.\par

If one needs to keep the second moment $\EL_2$ in the expression for $\beta(g_0)$ in Eq.~(\ref{Eq:betag0}), we get
\begin{equation}\label{Eq:betag0Exact}
 \beta(g_0)=-g_0^2-(a_0+a_1)g_0+c_0b_1-a_0a_1+c_0c_1\frac{\EL_2}{\EL_0}.
\end{equation}
For the stable late time fixed point $g_0=-2$, we get the value ${\EL_2}/{\EL_0}=1$ as expected. For the unstable late time fixed point 
$g_0=-3$, ${\EL_2}/{\EL_0}={3}/{8}$. The correction to $\beta(g_0)$ because of the term $\EL_2$ is also shown in Fig.~(\ref{fig:FS}b).\par

It is interesting to note here that the late time behavior of the ratio of moments for the unstable fixed point $g_n=-3$ (for any arbitrary $n$) turns
out to be
\begin{equation}\label{Eq:UnRatio}
 \lim_{\rho\rightarrow\infty}\frac{\EL_n(\rho)}{\EL_0(\rho)}=P_{2n}(0).
\end{equation}
An important observation from Eqs.~(\ref{Eq:UnRatio},\ref{Eq:StableFS}), is that the fixed points $g_n=-3$ and $g_n=-2$ correspond to vanishing effective longitudinal and transverse pressure $\hat{\PL}=0,\hat{\PT}=0$ respectively.
One notices that the late time behaviour of the ratio of moments for stable and unstable fixed points in the Bjorken scenario \cite{Blaizot:2017ucy} is exactly the opposite of Gubser
flow. This can be understood from the previous observation i.e., for Gubser flow  $\hat{\PT}\rightarrow 0$,  the system free streams in the transverse direction at late-times while for Bjorken flow $\hat{\PL}\rightarrow0$,  the system free-streams along the longitudinal direction.

\section*{The hydrodynamic fixed point}

Previously, we have seen that for an initially isotropic equilibrium distribution, the collisionless FS solution drives the system
to anisotropic distribution with a wealth of moments being generated as time goes on. Owing, to the presence of fixed point that we have already seen, 
the time dependence can be qualitatively captured by the two lowest moments.\par

As we have discussed earlier, the effect of collision is to wash out the effect of initial anisotropy and the anisotropy generated due to expansion. 
For a 1+1 dimensional expansion, the late time behaviour of the system can be described as a series expansion in the power of expansion scalar $1/\tau$. The late-time behaviour in such a system has been attributed to the presence of a different kind of the fixed point, 
called hydrodynamic fixed point. To proceed, we shall assume that all the moments $\EL_n$ admit a gradient expansion in powers of $\tanh\rho$ \cite{Denicol:2018pak}:
\begin{equation}\label{Eq:GE}
 \EL_n(\rho)=\sum_{m=n}^{\infty}{(\tanh \rho)}^m\gamma^{(m)}_n,
\end{equation}
where \(\gamma^{(m)}_{n}\)'s are coefficients of the expansion.
Due to Landau matching condition, for all orders, the zeroth moment corresponds to the energy density  i.e., $\gamma^{(m)}_0=\hat{\epsilon}$ for all $m$. Since $\hat{\epsilon}(\rho)$ is time-dependent, we deduce that coefficients $\gamma^{(m)}_n$ are also time-dependent. The late-time behaviour of the term $\gamma^{(0)}_0$ can be obtained from Eq.~(\ref{Eq:Main}) by ignoring the contribution from $\EL_1$. We find that the ideal hydrodynamics limit of the time evolution of $\hat{\epsilon}(\rho)$ is governed by the term $a_0=8/3$, i.e.,
$\hat{\epsilon}(\rho)\sim{(\cosh\rho)}^{-8/3}$. The leading terms in the expansion can be determined by demanding the cancellation of relaxation terms in Eq.~(\ref{Eq:Main}) since other terms can be ignored at large times. The above condition can be met if we have 
\begin{equation}\label{Eq:Leading}
\EL_{n}=\frac{-cb_n\tanh\rho}{\hat{T}(\rho)}\EL_{n-1}.
\end{equation}
Starting from $n=1$, the above equation leads to a recursion relation for $\EL_n$ from which the coefficients $\gamma^{(n)}_n$ can be deduced as
\begin{equation}
\gamma^{(n)}_n=\hat{\epsilon}{\left(\frac{c}{\hat{T}}\right)}^n(-1)^{n}\prod_{i=1}^nb_i.
\end{equation}
\begin{figure*}
    \begin{center}
    \includegraphics[width =0.45\linewidth]{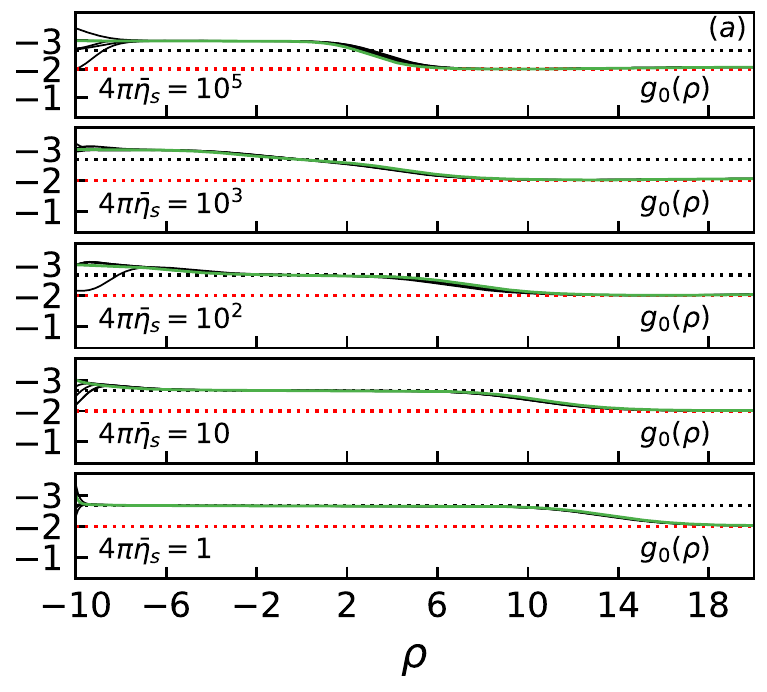}
     \includegraphics[width =0.46\linewidth]{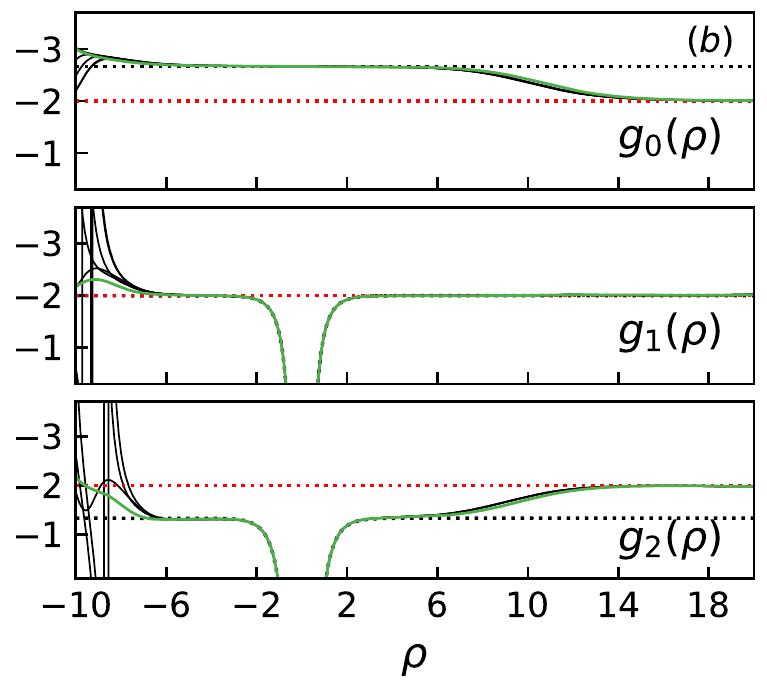}
    \end{center}
     \caption{(a) (Color online) Attractor solutions for $\mathcal{L}_0$ moments in terms of $g_0(\rho)$  truncating Eq.~(\ref{Eq:Main}) after 150 terms. Thin black solid lines correspond to arbitrary initial conditions, red dotted line corresponds to FS stable fixed point for $\rho>0$, i.e, $g_0=-2$, black dotted line corresponds to hydrodynamic fixed point $g_0=-8/3$. Green solid is defined as the  'attractor' solution. For all curves $\rho_0=-10$ with decreasing values of $\bar{\eta}_s$. (b) $g_0(\rho)-g_2(\rho)$ as a function of $\rho$ with intital condition fixed at $\rho=-10$ and $4\pi\bar{\eta}_s=10$. Line symbols are same as that of (a), except that black dotted lines are determined from Eq.~(\ref{Eq:HydroFP})}.
\label{Fig:Allg0}
\end{figure*}
\begin{figure}
\centering
\includegraphics[width=0.75\linewidth]{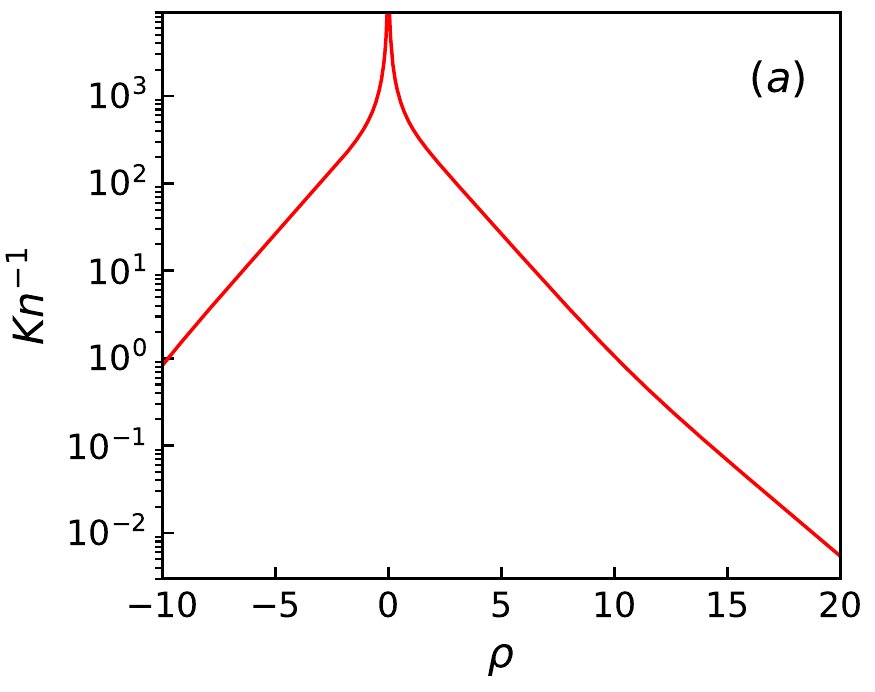}
\includegraphics[width=0.75\linewidth]{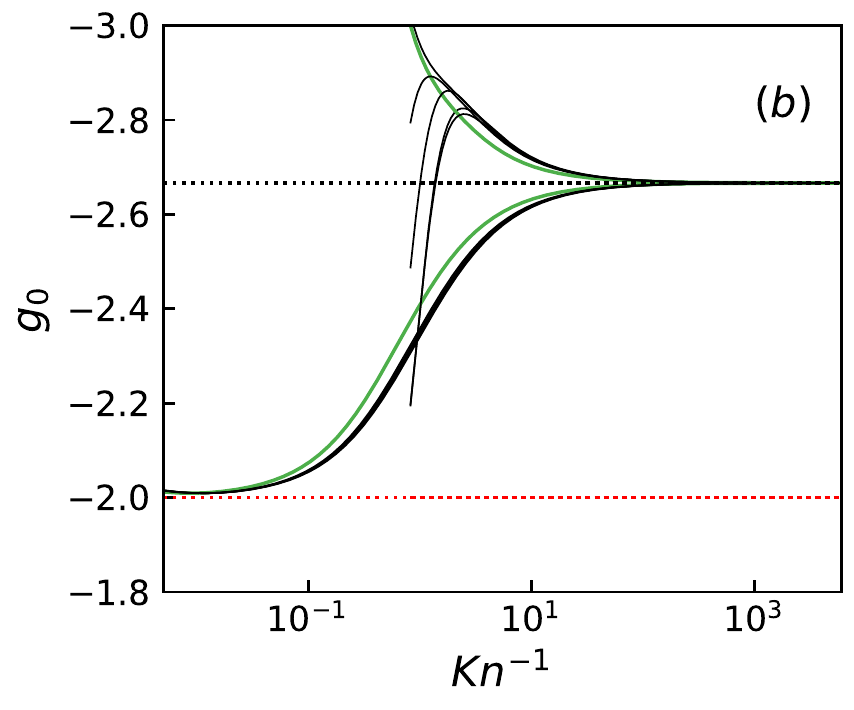}
\caption{(a) Inverse Knudsen number $\Kn^{-1}={(\hat{\tau}_R \lvert\tanh\rho\rvert )}^{-1}$ for Gubser flow, with initial conditions pertaining to $g_n=-3$ at $\rho=-10$, i,e, 'attractor' solution and $4\pi\bar{\eta}_s=10$. (b) $g_0$ as function of $\Kn^{-1}$ with the parameters set as that of (a). The line symbols are same as that of Fig.~(\ref{Fig:Allg0}).}
\label{Fig:InVKn}
\end{figure}
Since, $\hat{\epsilon}(\rho)\sim\hat{T}^4(\rho)$, the time dependence of leading order is given as: 
$\gamma_n^{(n)}\sim {(\cosh\rho)}^{{-2 (4-n)}/{3}}$. Substituting, this in Eq.~(\ref{Eq:GE})
gives us the time dependence of $\EL_n\sim{(\tanh\rho)}^n{(\cosh\rho)}^{{-2 (4-n)}/{3}}$. 
The fixed points can be calculated using the definition of $g_n(\rho)$ in Eq.~(\ref{Eq:gnDef}),
\begin{equation}\label{Eq:HydroFP}
 g_n(\rho)=\frac{2 (n-4)}{3}+n\csch^2(\rho).
\end{equation}
Eq.~(\ref{Eq:HydroFP}) shows that unlike the 1+1d Bjorken case where the $g_n(\rho)$ is a constant, for 2+1d Gubser flow the hydrodynamic fixed point is a function of time.
However, in the asymptotic limit, we have $g_n(\rho\rightarrow \pm\infty)={2 (n-4)}/{3}$.
 It is also interesting to note that for $n=1$ at large time, the hydrodynamic and FS
  stable fixed point are exactly same.\par
The next to leading order correction to $\gamma^{(n)}_n$ can be determined by ignoring the contribution from the term $\EL_{n+1}/\EL_n$ but still keeping the constant contribution in Eq.~(\ref{Eq:gnDef}). This gives
\begin{equation}
 \frac{\gamma^{(n+1)}_n}{\gamma^{(n)}_n}=-\frac{c}{\hat{T}}\left(\frac{2 (n-4)}{3}+a_n+b_n\frac{\gamma^{(n)}_{n-1}}{\gamma^{(n)}_n}\right).
\end{equation}

\section*{The hydrodynamic attractor}
Following \cite{Blaizot:2017ucy}, we define the 'attractor' as the solution of Eq.~(\ref{Eq:Main}) with initial conditions pertaining to the FS stable fixed point for $\rho<0$, i.e. $g_n=-3$ with the corresponding moments as given by Eq.~(\ref{Eq:UnRatio}). In Fig.~(\ref{Fig:Allg0}a), the numerical solution of attractor and arbitrary initial conditions is shown for $\EL_0$ in terms of the quantity $g_0$. While solving numerically we used $\rho_0=-10$, because this value avoids unphysical behaviour e.g. negative temperatures \cite{Denicol:2014tha} if one initialise at \(\rho_{0}=0\). The solution is obtained, for various values of $\bar{\eta}_s$ in decreasing order as one goes from top to bottom of the figure. The solution is obtained  after truncating Eq.~(\ref{Eq:Main})  after 150 terms. Thin black solid lines correspond to arbitrary initial conditions, red dotted line corresponds to FS stable fixed point for $\rho>0$, i.e, $g_0=-2$, black dotted line corresponds to hydrodynamic fixed point $g_0=-8/3$ (Eq.~(\ref{Eq:HydroFP}), with $n=0$) and the green solid line is the attractor solution respectively. We can make the following observations from Fig.~(\ref{Fig:Allg0}a)
\begin{enumerate}
 \item Starting from the top, with a large value of $4\pi\bar{\eta}_s=10^5$, which almost correspond to the FS limit in this setup, one finds that there are two plateaus, for early time this corresponds the FS stable fixed point $g_0=-3$ and at late times $g_0=-2$. Arbitrary initial conditions approach these FS attractor at early and late time. The system does not approach the hydrodynamic fixed point at any point of time as expected.
 \item Decreasing the value of $4\pi\bar{\eta}_s=10^3$, pushes the $g_0=-2$ fixed point further in $+\rho$ and $g_0=-3$ backward in $-\rho$. Here, the system just transiently touches the hydrodynamic fixed point around $\rho=0$ before approaching the late time FS fixed point $g_0=-2$. The same trend continues for $4\pi\bar{\eta}_s=10^2$, with the system staying relatively longer in the hydrodynamic fixed point than the previous value and a plateau developing.
 \item For $4\pi\bar{\eta}_s=10$, the system pushes the $g_0=-3$ fixed point far back in $-\rho$ and the hydrodynamic fixed point is approached relatively quicker than previous cases. At late times the hydrodynamic fixed point decays to the FS fixed point $g_0=-2$. The hydrodynamic attractor goes from $g_0=-3 \rightarrow -8/3 \rightarrow -2$. Further decrease of $4\pi\bar{\eta}_s=1$  plateaus the hydrodynamic fixed point longer in time before eventually decaying to $g_0=-2$.
\end{enumerate}
A comparison between 1+1d Bjorken expansion to the intrinsically 3+1d Gubser expansion is apt here. Unlike Bjorken flow which thermalizes/hydrodynamizes at a late time, Gubser flow goes from FS$\rightarrow$thermalizes/hydrodynamizes$\rightarrow$FS. This is because the inverse Knudsen number $\Kn^{-1}=\tau/\tau_R$ for Bjorken flow in the conformal limit grows with time \cite{Chattopadhyay:2019jqj}, in the other hand $\Kn^{-1}={(\hat{\tau}_R \lvert\tanh\rho\rvert )}^{-1}$ for Gubser flow increases for $\rho<0$ and decreases $\rho>0$ as shown in Fig.~(\ref{Fig:InVKn}a). We also plot the function $g_0$ as a function of inverse Knudsen number $\Kn^{-1}$ in Fig.~(\ref{Fig:InVKn}b), with $4\pi\bar{\eta}_s=10$. The arrow of time in this figure first moves to the right and then to left as can be understood from Fig.~(\ref{Fig:InVKn}a).  The initial condition of the setup corresponds to $\Kn^{-1}\sim 1$. The system moves to the right i.e. with increasing value of $g_0$ with increasing  $\Kn^{-1}$, reaches the hydrodynamic fixed point at few orders in $\Kn^{-1}$, turns left again, remaining in the hydrodynamic regime with decreasing value of $\Kn^{-1}$. After reaching $\Kn^{-1}<10^1$, the $g_0$ increases further until it reaches the FS fixed point at small values of $\Kn^{-1}$. Such a case was recently studied in \cite{Chattopadhyay:2019jqj} by varying the Knudsen number for a system undergoing Bjorken flow.\par

Next, we examine the behaviour of higher-order $g_n$, where $n=1,2$ as a function of $\rho$ Fig.~(\ref{Fig:Allg0}b). Following are the salient features: a) All curves approach the hydrodynamic fixed point around $\rho=0$  ($\Kn^{-1}\gg1$), which can be called the 'hydrodynamic regime'. As noted earlier all higher-order moments have a pole at $\rho=0$. b) The hydrodynamic and late time FS fixed point is the same for $n=1$. 
 c) We have checked that the hydrodynamic fixed-point values are independent of the truncation order but not the FS values, i.e., as one goes to higher-order in the set of Eqs.~(\ref{Eq:Main}), the match to late time exact values of FS fixed point $g_n=-2$  gets better.

\section*{Matching to hydrodynamics}
Here we pause and compare the various orders of moments generated from truncation of Eq.~(\ref{Eq:Main}) to successive orders of viscous hydrodynamic corrections. In the last
section we have already shown that truncation at lowest order with vanishing $\EL_1$ results in ideal hydrodynamic equation of motion. Truncation at $n=1$ yields two coupled set of equations:
 \begin{align}
  \frac{\partial \hat{\epsilon}}{\partial \rho}&=-\tanh(\rho)\left(\frac{8}{3}\hat{\epsilon}-\hat{\pi}\right),\nonumber\\
  \frac{\partial\hat{\pi}}{\partial\rho}&=-\tanh (\rho ) \left(\frac{2}{3}b_1\hat{\epsilon}+a_1\hat{\pi}\right)-\frac{\hat{\pi}
   \hat{T}}{c},\label{Eq:secondorder}
 \end{align}
where $\hat{\pi}=-c_0\EL_1$. The above set of equations constitute the second order viscous hydrodynamic equations for Gubser flow in the Chapman-Enskog expansion \cite{Chattopadhyay:2018apf}. The term $b_1$, $a_1$ are related to the the term $\hat{\beta}_{\pi}$, $\hat{\lambda}_{\pi}$ through the relation: $\hat{\beta}_{\pi}=-b_1\hat{\epsilon}/2$, $\hat{\lambda}_{\pi}=a_1$ respectively and as we have already mentioned $\hat{\tau}_R=c/\hat{T}(\rho)$. In the first order Navier-Stokes approximation the Eq.(\ref{Eq:secondorder}) reduces for small $c$: $\hat{\pi}=\left(4/3\right)\hat{\tau}_R\hat{\beta}_{\pi}\tanh\rho$.

For third and higher-order viscous correction from the moment equation by keeping the term $\EL_2$ in Eq.~(\ref{Eq:secondorder}) and also considering the time evolution of moment $\EL_2$:
 \begin{align}
  \frac{\partial\hat{\pi}}{\partial\rho}&=-\tanh (\rho ) \left(\frac{2}{3}b_1\hat{\epsilon}+a_1\hat{\pi}+\frac{2}{3}c_1\EL_2\right)-\frac{\hat{\pi}
   \hat{T}}{c},\\
  \frac{\partial{\EL}_2}{\partial\rho}&= -\tanh (\rho ) \left(a_2\EL_2+\frac{3 }{2}b_2\hat{\pi}\right)-\frac{\EL_2\hat{T}}{c}.
 \end{align}

In the above equation, $\EL_2$ appears as a new dynamical variable with its own evolution equation which in contrast to third-order hydrodynamics is related to $\hat{\pi}$ by a constitutive relation. To make a connection with hydrodynamics it suffices to take a small $c$ limit in which,
\begin{equation}\label{Eq:ThirdOrder}
 \EL_2=-\frac{3}{2}\hat{\tau}_Rb_2\hat{\pi}\tanh\rho=-\frac{9b_2}{8\hat{\beta}_\pi}\hat{\pi}^2,
\end{equation}
where we have used the Navier-Stokes limit to express $\tanh\rho$ in terms of moment $\hat{\pi}$. Eq.~(\ref{Eq:ThirdOrder}) can be matched to the coefficient $\hat{\chi}$ appearing in third-order hydrodynamic \cite{Chattopadhyay:2018apf} through the relation: $\hat{\chi}=-(3/4)c_1b_2$. For higher-order corrections, the series $\gamma_n^{(n)}$ is divergent since it grows as $n!$ for large $n$ as has already been seen previously in Bjorken flow.
\section*{Conclusion}
To summarize, the moment method which has been formulated in \cite{Blaizot:2017ucy} for 1+1d boost invariant system acts
both as a practical tool for solving the kinetic equation and to address dynamical questions like the presence of fixed points
and attractors, etc. for the system under consideration. In the present work, we applied the moment method for a system undergoing
Gubser flow which has a simultaneous transverse and longitudinal expansion. Our study suggests that the presence of an attractor,
to which the solution of the dynamical equations quickly converges before eventually reaching the viscous hydrodynamic regime is
a feature not limited to the 1+1d system with an overwhelming amount of symmetries like Bjorken flow. However, unlike Bjorken flow which starts with vanishing longitudinal pressure at an early time and ends up in thermalization at late times, the dynamics of Gubser flow is different. Initially, the dynamics of the system is similar to Bjorken flow but ends up with vanishing transverse pressure, with thermalization happening in an intermediate stage. We also compared the numerical solution obtained after solving the coupled moment equations to the exact solution which shows a very good agreement. A similar comparison of coefficients obtained through this method with successive orders of viscous hydrodynamic corrections shows exact agreement.  We believe that these results offer detailed insights into the dynamics of longitudinal/transverse momentum isotropization in relativistic systems undergoing simultaneous transverse and longitudinal expansion. In passing, we would also mention that although Gubser flow has a transverse expansion, in itself is a highly idealized model when confronted with dynamics of matter produced in heavy-ion collisions. Therefore we think it is important to investigate the appearance of attractors further, by relaxing certain symmetries e.g. conformality and homogeneity along lines similar to \cite{Romatschke:2017acs} but with more analytical control e.g. by adding mass terms or using a non-trivial metric choice.
\section*{Acknowledgements}
AD thanks Ananta Prasad Mishra, Amaresh Jaiswal and Sunil Jaiswal for fruitful discussions. AD also wishes to thank Ulrich Heinz and Chandrodoy Chattopadhyay for their useful comments which clarified certain disjoint pieces of this work. AD and VR acknowledge financial support from DAE, Government of India. VR is also supported by the DST INSPIRE Faculty research grant, India.

\bibliographystyle{elsarticle-num-names}
\bibliography{sample.bib}







\end{document}